\documentclass{ws-procs9x6_mod}
\usepackage[super,nospace]{cite}

\begin{document}

\title{Constraints on Dissipative Non-Equilibrium Dark Energy Models from
Recent \\ Supernova Data}

\author{V.~A.~MITSOU}

\address{Instituto de F\'{i}sica Corpuscular (IFIC),
CSIC -- Universitat de Val\`{e}ncia, \\
Edificio Institutos de Investigaci\'{o}n, P.O.\ Box 22085, E-46071 Valencia, Spain  \\
E-mail: vasiliki.mitsou@cern.ch}

\maketitle

\abstracts{ Non-critical string cosmologies may be viewed as the analogue of
off-equilibrium models arising within string theory as a result of a cosmically
catastrophic event in the early Universe. Such models entail relaxing-to-zero
dark energies provided by a rolling dilaton field at late times. We discuss
fits of such non-critical models to high-redshift supernovae data, including
the recent ones by HST and ESSENCE and compare the results with those of a
conventional model with Cold Dark Matter and a cosmological constant and a
model invoking super-horizon perturbations. }

\section{Introduction}

There is a plethora of astrophysical evidence today, from supernovae
measurements,\cite{riess,SNLS,HST,essence,davis} the cosmic microwave
background\cite{wmap}, baryon oscillations\cite{baryon} and other cosmological
data, indicating that the expansion of the Universe is currently accelerating.
The energy budget of the Universe seems to be dominated at the present epoch by
a mysterious dark energy component. Many theoretical models provide possible
explanations for the latter, ranging from a cosmological
constant\cite{concordance} to super-horizon perturbations\cite{riotto} and
time-varying quintessence scenarios,\cite{steinhardt} in which the dark energy
is due to a smoothly varying scalar field dominating cosmology in the present
era. In the context of string theory, such a time-dependent `quintessence'
field is provided by the scalar dilaton field of the gravitational string
multiplet\cite{aben,gasperini,emnw}.

\section{Dissipative Q-Cosmology Basics}

Most of the astrophysical analyses so far are based on effective
four-dimensional Robertson-Walker Universes, satisfying on-shell dynamical
equations of motion of the Einstein-Friedman form. Even in modern approaches to
brane cosmology, described by equations deviating during early eras of the
Universe from the standard Friedman equation, the underlying dynamics is
assumed to be of classical equilibrium (on-shell) nature.

However, cosmology may not be an entirely classical equilibrium situation.  The
initial Big Bang or other catastrophic cosmic event, such as a collision of two
brane worlds in the modern approach to strings, which led to the initial rapid
expansion of the Universe, may have caused a significant departure from
classical equilibrium dynamics in the early Universe, whose signatures may
still be present at later epochs including the present era. One specific model
for the cosmological dark energy which is of this type, being associated with a
rolling dilaton field that is a remnant of this non-equilibrium phase, was
formulated\cite{emnw,diamandis} in the framework of non-critical string
theory\cite{aben,emn}. This scenario is called `Q-cosmology'. It is of outmost
importance to confront the currently available precision astrophysical data
with such non-equilibrium stringy cosmologies. The central purpose of this talk
is to present a first step towards this direction, namely a confrontation of
cosmological data on high-redshift supernovae\cite{EMMN,MM} with Q-cosmologies
and compare the results with the predictions of the conventional $\Lambda$CDM
model\cite{concordance} and the super-horizon model.\cite{riotto} Care must be
taken in interpreting the Q-cosmology scenario. Since such a non-equilibrium,
non-classical theory \emph{is not described by the equations of motion derived
by extremising an effective space-time Lagrangian}, one must use a more general
formalism\cite{EMMN} to make predictions that can be confronted with the
current data.

\section{Supernova Data Analysis}

We use recent type-Ia supernovae (SN) data released by the Hubble Space
Telescope (HST)\cite{HST} and the ESSENCE collaboration.\cite{essence} Among
16~newly discovered high-redshift SNe,\cite{HST} the so-called `gold' dataset
\emph{(Riess07)} embraces SNe from other sets: 14~SNe discovered earlier by
HST,\cite{riess} 47~SNe reported by SNLS,\cite{SNLS} and 105~SNe detected by
ground-based discoveries, amounting a total of 182 data points. An additional
set of 77~SNe tagged as `silver' due to lower quality of photometric and
spectroscopic record is also listed. The ESSENCE dataset \emph{(WV07),\/} on
the other hand, consists out of 60~SNe of $0.015<z<1.02$ discovered by
ESSENCE,\cite{essence} 57~high-$z$ SNe discovered during the first year of
SNLS\cite{SNLS} and 45~nearby SNe. The results presented here were obtained by
analysing a compilation of the aforementioned data
\emph{(WV07+Riess07)},\cite{davis} normalised to account for the different
light-curve-fitters employed.

These data are given in terms of the distance modulus $\mu = 5\log d_L + 25$,
where the luminosity distance $d_L$ (in megaparsecs) for a flat universe is
related to the redshift $z$ via the Hubble rate $H$: $d_L = c(1 + z) \int_0^z
\frac{dz'}{H(z')}$. We note that this observable depends on the expansion
history of the Universe from $z$ to the present epoch, and recall that,
although most of the available supernovae have $z < 1$, there is a handful with
larger values of $z$. In the analysis that follows, the predictions for the
Hubble rate $H(z)$ of the following three cosmological models are investigated:

\paragraph{\boldmath$\Lambda$CDM:\unboldmath}
  In a CDM model with a cosmological constant,\cite{concordance} we have
  \begin{equation}
H(z) = H_0 \left(\Omega_{\rm M} ( 1 + z)^3 + \Omega_\Lambda ( 1 + z )^{3(1 +
w_0)}\right)^{1/2}.\label{eq:lcdm}
\end{equation}

\paragraph{Super-horizon model:} The Universe is assumed to
  be filled with non-relativistic matter only,\cite{riotto} and there is no dark energy of
  any sort:
  \begin{equation}
H(z) = {\overline a}^{-1}\frac{d{\overline a}(t)}{dt}= \frac{{\overline H}_0}{1
- \Psi_{\ell 0}}\left(a^{-3/2} - a^{-1/2}\Psi_{\ell 0}\right), \label{eq:shcdm}
\end{equation}
where $(1 + z)^{-1} = \overline{a}(t)$, $\Psi(\vec{x},t)$ is the gravitational
potential, and $\Psi_{\ell 0}$ is a free parameter.

\paragraph{Q-cosmology:} A parametrisation for $H(z)$ in the Q-cosmology
framework at late eras, such as the ones pertinent to the supernova and other
data ($0<z<2$), where some analytic approximations are allowed\cite{EMMN}, is
used in the analysis:
\begin{equation}\label{eq:formulaforfit}
\frac{H(z)}{H_0} = \sqrt{{\Omega }_3 (1 + z)^3 + {\Omega }_{\delta} (1 +
z)^\delta + {\Omega}_2(1 + z)^2}~,~~{\Omega }_3 + {\Omega}_{\delta} +
{\Omega}_2 = 1,
\end{equation}
with the densities $\Omega_{2,3,\delta}$ corresponding to present-day values
($z = 0$). However, a complete analysis of the non-critical and dilaton
effects, which turn out to be important in the present era after the inclusion
of matter, requires a numerical treatment.\cite{diamandis} In general, the
three parameters to be determined by the fit are $\Omega_3$, $\Omega_\delta$
and $\delta$. Here, a fixed value of $\delta=4$ is assumed for simplicity,
justified by earlier results,\cite{EMMN} whilst a more complete analysis is
given in Ref.~\refcite{MM}.

\begin{figure}[ht]
\centerline{\epsfxsize=3.7in\epsfbox{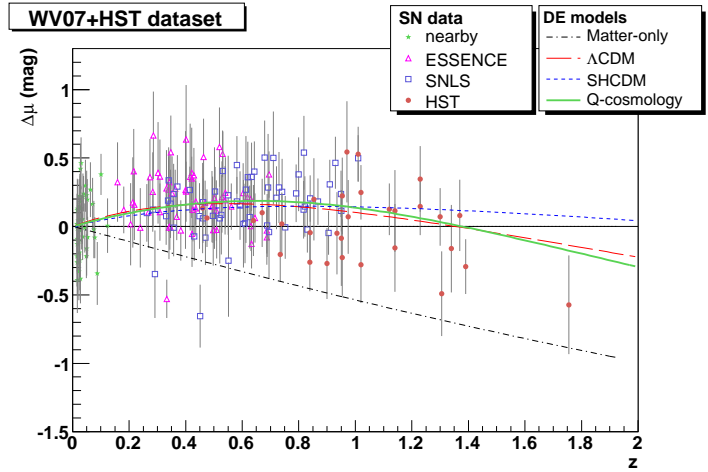}} \caption{Residual
magnitude versus redshift for supernovae from the WV07+HST dataset and model
predictions for the best-fit parameter values.\label{figure}}
\end{figure}

For illustration purposes, both data and predictions of cosmological models are
expressed in the following as residuals, $\Delta\mu$, from the empty-Universe
prediction (Milne's model, $\Omega_{\rm M}=0$). The VW07+Riess07 dataset, which
amounts to a sample of 192~supernovae in total, is shown in Fig.~\ref{figure},
where the predictions of the cosmological models under study are also displayed
for the best-fit parameter values, listed in Table~\ref{table}.

The analysis involves minimisation of the standard $\chi^2$ function with
respect to the cosmological model parameters. The best-fit parameter values,
the $1\sigma$ errors and the corresponding $\chi^2$ values are listed in
Table~\ref{table} for the three cosmological models.
\begin{table}[ph] \tbl{Fit
results for the WV07+HST dataset for various models.} {\footnotesize
\begin{tabular}{@{}lccc@{}}
\hline
{} &{} &{} &{}\\[-1.5ex]
Model & Best-fit parameters & $\chi^2$ & $\chi^2/{\rm dof}$\\[1ex]
\hline
{} &{} &{} &{}\\[-1.5ex]
$\Lambda$CDM flat & $\Omega_{\rm M}=0.259\pm0.019$ & 196 & 1.02\\[1ex]
$\Lambda$CDM      & $(\Omega_{\rm M},\Omega_\Lambda$)=(0.33,\,0.85) & 195 & 1.03 \\[1ex]
Super-horizon     & $\Psi_{\ell0}=-0.90\pm0.07$ & 200 & 1.05\\[1ex]
Q-cosmology       & $\Omega_3=-2.8\pm0.5$, $\Omega_4=0.86\pm0.22$ & 195 & 1.02\\[1ex]
\hline
\end{tabular}\label{table} }
\vspace*{-13pt}
\end{table}

It is evident from Fig.~\ref{figure} and Table~\ref{table} that the standard
$\Lambda$CDM model fits the supernova data \emph{very well}, as expected from
earlier analyses. The super-horizon dark matter model also fits the supernova
data \emph{quite well}. Both of these models are {\it on-shell}, i.e., they
satisfy the pertinent Einstein's equations. Moreover, {\it off-shell} cosmology
models are also compatible with the data. As we discussed above, off-shell
effects are important in our Q-cosmology model. Introducing the appropriate
parametrisation Eq.~(\ref{eq:formulaforfit}) to allow for these off-shell
effects, we find that \emph{the Q-cosmology model may fit the supernovae data
as well as the standard $\varLambda$CDM model}. Further constraints\cite{MM} on
cosmological models may be imposed by observations of high-$z$ red galaxies
constraining the Hubble parameter, $H(z)$.\cite{Hz}

\section{Conclusions}

We show that recent high-redshift supernovae data confirm\cite{MM} the
constraints established in earlier studies\cite{EMMN} which demonstrated that
cosmological models with no dark energy may be viable alternatives to the
Standard $\Lambda$CDM model. As more precision astrophysical data are coming
into play, more stringent constraints can be imposed on our non-critical string
Q-cosmologies.

\end{document}